\documentclass[12pt]{iopart}
\usepackage{dcolumn}
\usepackage{epsfig}

\usepackage{graphicx}
\usepackage{longtable}
\usepackage{dcolumn}
\usepackage{bm}

\def\etal{{\em et al. }}

\def\CSIX{C$_{60}\,$}
\def\CSEV{C$_{70}\,$}
\def\afz{$\alpha_0$}
\def\avib{$\alpha_{vib}\,$}
\def\amean{$\alpha_{el}\,$}
\def\ael{$\alpha_{el}\,$}

\def\aten{$\alpha_{ij}\,$}

\def\AAA{\AA$^{3}$ }

\begin{document}

\title[  Static dipole polarizability of C$_{70}$ fullerene ]{
 Static dipole polarizability of C$_{70}$ fullerene }

\author{Rajendra R. Zope}

\address{Department of Physics, University of Texas at El Paso, El Paso 79968}
\ead{rzope@utep.edu}
\date{\today}

\begin{abstract}
    The electronic and vibrational contributions to the 
static dipole polarizability of C$_{70}$ fullerene are determined 
using the {\em finite-field } method within the density functional formalism.
Large polarized Gaussian basis sets augmented with diffuse functions are used
and the exchange-correlation effects are described within the Perdew-Burke-Ernzerhof
generalized gradient approximation (PBE-GGA).  The calculated polarizability ($
\alpha$) of C$_{70}$ 
is 103 \AAA, in excellent agreement with the experimental value of 102 \AAA\, and is 
completely determined by the electronic part, vibrational contribution being
negligible.  The ratio  $\alpha(C_{70})/\alpha( C_{60})$ is 1.26. 
The comparison of polarizability calculated with only local terms (LDA)
in the PBE functional to that obtained with PBE-GGA shows that LDA
is sufficient to determine the static dipole polarizability of C$_{70}$.
\end{abstract}

\pacs{ }

\submitto{\jpb}

\maketitle
    
\section{Introduction}
  C$_{70}$  is perhaps the most studied carbon fullerene after the C$_{60}$ fullerene. 
Several studies have addressed electric response properties 
of  C$_{70}$  and its more famous cousin  C$_{60}$
\cite{C70:378, C70:377, C70:375, C70:382, C70:379, C70:385, C70:386, C70:387, C70:393, C70:390, C70:391,
C70:381, C70:389, C70:383, C70:376, C70:374, C70:388}. 
Static dipole polarizability  \afz\, 
of  C$_{70}$, which is the subject of present work,  was recently measured 
in gas phase by  Compagnon and coworkers\cite{C70:375}.
Using the  molecular beam deflection technique, they reported the mean static 
dipole polarizability to be  102 \AA$^3$\,  with an error bar of $\pm 14$ \AA$^3$. 
  Theoretically, polarizability  of C$_{70}$ has been subject of several 
investigations\cite{C70:382, C70:379, C70:385, C70:393, 
 C70:376, C70:374}.
Only two of these studies are, however, at the {\it ab initio} level of theory.
The first study is by  Jonsson and coworkers\cite{C70:379}, who using 
the self-consistent-field (SCF), and the multi-configuration
self-consistent field  (MCSCF) theories in combination 
with the 6-31++G basis set reported polarizability of  C$_{70}$  to be 89.8 \AA$^3$. 
The second study is due to van Fassen \etal\cite{C70:374} who
used time dependent density functional theory (TDDFT) and three  
different basis sets to calculate \afz\, of \CSEV \cite{C70:374}. Using the largest 
basis (triple zeta with additional field induced polarization function TZVP+)
they found \afz\, to be 104.8 \AA$^3$.  These authors also computed \afz\, using the 
current-dependent Vignale-Kohn functional and concluded that polarizabilities 
calculated using the current-dependent functional 
gives good agreement with experimental values for \CSIX\, and  C$_{70}$ . The 
geometric structure of C$_{70}$  was not optimized in both these works.

    The present article complements these earlier studies and reports the 
static dipole polarizability of C$_{70}$ calculated  within 
density functional formalism, using large polarized Gaussian basis sets 
augmented by diffuse functions.  The calculations are performed  within 
the generalized gradient approximation using 
Perdew-Burke-Ernzerhof  parametrization\cite{PBE}.
Unlike in previous works, we first determine the equilibrium structure of the 
C$_{70}$ fullerene at the same level of theory. We then compute the 
static dipole polarizability.  We investigate
the vibrational  \avib\, contribution  to the polarizability \afz\
as well as the electronic  \ael\, contribution.
The former, which has not yet been computed,  
is computationally significantly more expensive than the latter
as it requires 
multiple optimization of C$_{70}$ geometry in the presence of electric 
field or requires the calculation of full vibrational spectrum.

\begin{figure}
\epsfig{file=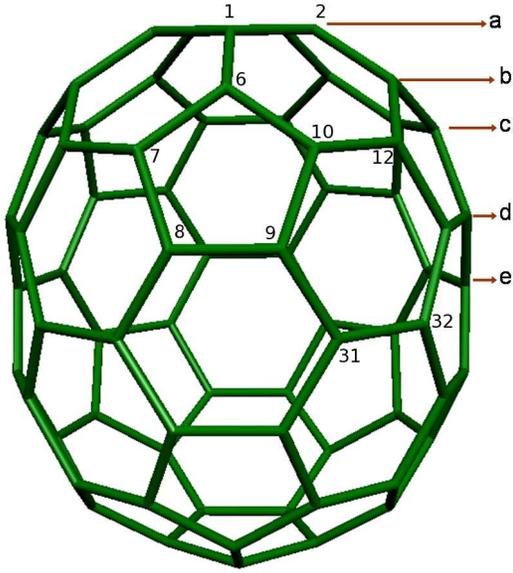,width=8.5cm,clip=true}
\caption{\label{fig:raman} The optimized geometry of C$_{70}$ fullerene. The 
five inequivalent atoms in the $D_{5h}$  structure are denoted by lower case letters.
The bond distance between labeled atoms are given in Table I.}
\end{figure}

\section{Computational details and Results}

 The geometry of C$_{70}$ was fully optimized using the limited memory 
Broyden-Fletcher-Goldfarb-Shanno algorithm\cite{lbfgs} using the NRLMOL suit of codes developed 
by Pederson and coworkers\cite{NRLMOL1,NRLMOL2,NRLMOL_Basis}.
The code employs an efficient scheme for numerical quadrature for the exchange-correlation 
integrals\cite{NRLMOL_MESH}. 
The $D_{5h}$ symmetry\cite{C70:380} of C$_{70}$ fullerene 
was exploited 
to reduce the computational expenses during structure optimization.
The molecular orbitals in NRLMOL are expressed as a linear combination 
of Gaussian orbitals. The Gaussian basis set for 
C consists of 5 $s-$, 3 $p-$, and 1 $d-$ type Gaussians each contracted 
from 12 primitive functions. The total number of basis functions used for the 
structure optimization is 2450 and that used for polarizability 
calculations is 2870.  The exponents in the basis set are optimized 
iteratively by performing a self-consistent calculation 
on isolated atoms\cite{NRLMOL_Basis}. 
More details about basis set and its construction can be 
found in Ref. \cite{NRLMOL_Basis,PBAS05,PB_POL}.
The fully optimized structure of \CSEV\,
is shown in Fig. 1.  There are five inequivalent atoms in C$_{70}$. 
The optimized positions of these inequivalent atoms in atomic units  are
  a   ( -2.3127,        0.0000,       -7.4352 ), 
  b   ( -4.5356,        0.0000,        6.0395 ), 
  c   (  5.5080,        1.3063,        4.5799 ), 
  d   (  5.9336,        2.6454,        2.2646 ), 
  e   (  6.5245,        1.3720,        0.0000 ).
The C$_{70}$ fullerene geometry can be generated using the $xyz$ coordinates
of inequivalent atoms and symmetry operations of point group $D_{5h}$, where 
the five-fold highest symmetry axis is the $z-$ axis. 
In Table I, the calculated bond distances are compared with some experimental and 
theoretical values reported in literature.  Agreement with experimental 
bond distances is quite good. The equatorial bond distance of 1.45 \AA\, 
is smaller than the 1.538 \AA\, measured in the gaseous electron diffraction experiment.
However, the smaller values comparable to the prediction of the present calculation, 
have been reported in other experiments.
The distance between the polar pentagons is 7.869 \AA\, in good agreement with 
experimental value of 7.906 \AA\, from the most recent experimental measurement\cite{C70:380}.

          The elements of the static dipole polarizability tensor are 
  \begin{equation}
         \alpha_{ij} = -\frac{\partial^2 E }{{\partial F_i } {\partial F_j}}.
             \label{eq:alpha_ij}
  \end{equation}
    Here, E is the total molecular energy and F$_i$ is the $i^{th}$ component of the 
 electric field. 
Eq. (\ref{eq:alpha_ij})  is the coefficient of the second term in  
 Taylor's expansion of total energy E in the presence of field:
  \begin{equation}
   E(F) = E_0 + \sum_i \Bigl ( \frac {\partial E}{\partial F_i} \Bigr ) F_i + 
          \frac{1}{2} \sum_{i,j}\Bigl ( \frac {\partial^2 E}{\partial F_i \partial F_j}\Bigr )  F_i  F_j
          + \cdots
           \label{eq:Taylor}
  \end{equation}
    Alternatively, the polarizability tensor elements
 could also be obtained from the induced dipole moments. 
 A number of methods have been formulated to  obtain \aten and several review
articles describing the 
details and applications  of these methods exist\cite{Bonin,Karna}. 
In this work we compute \aten  numerically using finite-difference formula. This 
requires calculation of total energy of the molecule for various field values.
The electronic contribution to the polarizability  \ael 
is  then obtained using a suitable approximation to compute the second derivative 
in  Eq. \ref{eq:alpha_ij}.   Alternatively, least-square fitting technique can also be 
used to extract the polarizability\cite{cohen:s34,Na_Pol,C70:390}. This way of calculating \ael 
is known as the {\em finite-field} (FF) method.  The implementation of FF method 
is fairly straight-forward.
The Hamiltonian of the system is augmented by the term $-\vec{F}\cdot\vec{r}$ that 
represent interaction of an electron with applied electric field $\vec{F}$. The 
self-consistent solution is performed to obtain total energy for  a given value 
of electric field. The self-consistent process takes into account the field-induced 
polarization or the so-called {\em screening} effects. The drawback is that 
addition of  $-\vec{F}\cdot\vec{r}$ to the Hamiltonian, in general, breaks 
the full point-group symmetry. Thus computational advantage of the 
full point group symmetry is lost.  In some cases, 
it may be still possible to use lower symmetry for applied fields along a
specific direction. The present calculations of \ael of \CSEV did not use any
symmetry. 
The electric field step was chosen by ensuring that the calculated 
polarizability is accurate and uncontaminated by higher polarizability. The \ael obtained 
from induced dipole moments agrees with that obtained from energy within 1\%.
The basis set used during 
structure optimization was augmented by a $d-$ type diffuse function
with 0.0772097 exponent.
This methodology has been found to provide a good description of \ael 
of C$_{60}$ and several molecules 
and clusters\cite{C70:390,C70:391,Jackson,PBAS05,PB_POL}.
For more 
details of methodology and application we refer reader to a recent review by 
Pederson and Baruah\cite{PB_POL}. The computational scheme used in this work has 
also been applied to \CSIX\, fullerene. In fact, one of the earliest 
calculation of  polarizability  of  \CSIX\,  by Pederson and Quong  used the 
same methodology\cite{C70:390,C70:391}. 
 As mentioned earlier, the nuclear positions were assumed to be 
frozen during the calculation of  \ael. The relaxation of nuclear positions 
in the presence of applied electric field also contribute to the 
polarizability. This contribution 
is often called vibrational polarizability \avib\, and  usually is the second 
largest contribution of the polarizability.   In some cases, particularly for 
the system with ionic or hydrogen bonding it could be even larger than \ael.
Here, we compute \avib within double-harmonic approximation. 
A full account of 
the formulation of calculation of \avib\, used in this work can be found 
in  Ref. \cite{PBAS05}.

          The mean or average polarizability  \amean, is one-third   of the 
trace of the polarizability matrix: 
 \begin{equation}
\alpha_m = \frac{1}{3} (\alpha_{xx} + \alpha_{yy} + \alpha_{zz}).
 \end{equation}
      The calculated values of the polarizability components  are $\alpha_{xx} = \alpha_{yy} =99$\AAA\,
and $\alpha_{zz} = 111$\AAA, where $z-$ axis is along the five-fold axis.
The mean polarizability \amean is 103 \AA$^3$ . 
We have also calculated unscreened polarizability using the sum-over-states expression 
with excitation energies approximated by eigenvalue differences. As in case of C$_{60}$ 
fullerene\cite{C70:390,C70:391}, the unscreened polarizability turns out to 
be 330 \AAA\,, roughly three times larger than the screened polarizability.
In Table II calculated \amean 
is compared with earlier published theoretical and  experimental  values.
  Recent measurements of gas-phase polarizability of \CSEV\,  indicate \amean 
to be 102 \AAA with a rather larger error bar of about 14 \AAA.  The predicted 
(PBE/NRLMOL) value
of  103 \AAA  agrees well with the experimental value. As can be seen from the Table 
II, all {\it ab-initio} values fall within the error bar. The PBE/NRLMOL \ael\, agree well 
with time dependent density functional theory calculation 
with statistically averaged orbital potential (TDDFT/SAOP) polarizability\cite{C70:374}. 
But, it is larger than the predictions  by Hartree-Fock (HF) theory 
(HF/6-31+G)\cite{jonsson:572} and 
time dependent current density functional theory (TDCDFT/VK)\cite{C70:374}models. 
The differences in the structure of C$_{70}$ used in those calculations and
that optimized in this work is one source of discrepancy. 
In particular, the higher order polarizabilities are  often sensitive to the molecular 
structure\cite{maroulis:583}.
The other possible cause of differences is in the modeling of many body effects.
In our recent study that compared  static dipole polarizabilities of 142 
small molecules in the HF and PBE models, the HF polarizabilities were found 
to be smaller than their PBE counterparts\cite{SOS}. 
The larger value of \amean in the PBE/NRLMOL 
than in the HF model is consistent with this observation. The  \amean\, obtained 
within the time-dependent current density functional model (TDCFT/VK)  
falls intermediate between the HF/6-31+G and PBE/NRLMOL predictions.   
We repeated calculations with only local terms in the PBE functional,
to estimate the effect of gradient correction to the exchange-correlation
functional. This approximation gives \amean\, to be 100 \AAA, 
indicating that local approximation is sufficient to determine
the dipole polarizability of C$_{70}$. 
The vibrational contribution to the polarizability tensor
within the double harmonic approximation\cite{PBAS05} is given 
as $\alpha_{vib,i,j} = \sum_{\mu} Z_{i,\mu} \omega^{-2}_{\mu} 
                    Z^{T}_{j,\mu}. $ Here, $\omega_{\mu}$ is the 
frequency of the $\mu$th vibrational mode, $Z_{i,\mu}$ is the effective 
charge tensor (See Ref. \cite{PBAS05} for details). 
  The vibrational 
contribution along the five-fold axis is $\alpha_{zz}^{vib}$  is 0.43 \AAA\,
while that along the transverse axis is 0.74 \AAA. Thus, in comparison with 
electronic polarizability,  vibrational 
contribution to the dipole polarizability of \CSEV is negligible and 
the static dipole polarizability of \CSEV\, is completely determined 
by the  electronic polarizability. This result in consistent with 
the finding of Pederson and coworkers\cite{PBAS05} for the \CSIX\, 
and is relevant for the dielectric response of carbon nanotubes.

    To conclude, static dipole polarizability of \CSEV is calculated 
within density functional formalism using large polarized Gaussian 
basis sets.  The calculated values of electronic
($\alpha_{el} = 102.8$ \AAA) and vibrational ($\alpha_{vib} = 4.3$ \AAA)
polarizabilities indicate that 
the vibrational contribution to the total polarizability of 
\CSEV is very small.

        This work is supported in part by 
the National Science Foundation through CREST grant,
by the University of Texas at El Paso and  by
the Office of Naval Research.
Authors acknowledge the computer time at the UTEP Cray 
acquired using ONR 05PR07548-00 grant.

\begin{table*}
\caption{The comparison of selected bond distances calculated in this work (PBE/NRLMOL) with 
those reported in literature. GED is the gaseuous electron diffraction measurements\cite{C70:380};
Solid-State Electron diffraction (SED)\cite{C70:394}; 
Neutron difftraction (ND) mesurments \cite{C70:384};
X-ray diffraction\cite{Xray}; Hartree-Fock/Dobule Zeta Basis \cite{Scuseria91}; BP86/TZP\cite{BP86};
PBE/NRLMOL (present).}
\begin{tabular}{lccccccc}
\hline
Bond distance &  GED  &  SED  &  ND  &  X-ray  &  SCF/DZP  &  BP86/TZP  & PBE/NRLMOL \\
\hline
C1-C2   &  1.461   &  1.464  &   1.460  &   1.458  &   1.451  &   1.454 & 1.439 \\
C1-C6   &  1.388   &  1.37  &   1.382  &   1.380  &   1.375  &   1.401  &  1.389 \\
C6-C7   &  1.453   &  1.47  &   1.449  &   1.459  &   1.446  &   1.450   &  1.436 \\
C10-C12   &  1.386   &  1.37  &   1.396  &   1.370  &   1.361  &   1.395   & 1.383 \\
C7-C8   &  1.468   &  1.46  &   1.464  &   1.460  &   1.457  &   1.449   &  1.433 \\
C8-C9   &  1.425   &  1.47  &   1.420  &   1.430  &   1.415  &   1.441   & 1.426  \\
C9-C31   &  1.405   &  1.39  &   1.415  &   1.407  &   1.407  &   1.424   & 1.410  \\
C31-C32   &  1.538   &  1.41  &   1.477  &   1.476  &   1.475  &   1.471   &  1.452 \\
\hline
\end{tabular}
\end{table*}


\begin{table*}
\caption{The comparison of calculated polarizability in (\AA$^3$) with 
the experimental and theoretical values reported in literature.}
\begin{tabular}{lccccccccc}
\hline
Method   &&   C$_{60}$   &&   C$_{70}$  &&   C$_{70}$/C$_{60}$  && Reference \\
\hline
Gas phase &&  76.5$\pm$8  && 102.$\pm$14  && 1.33  &&  Ref. \cite{C70:375}  \\
Ellipsometry    &&   79.0        &&   97.0       &&  1.23 && Ref. \cite{Ren:124} \\
EELS &&   83.0       &&  103.5        && 1.25  && Ref. \cite{Sohmen}  \\
{\bf Theory}   && && && \\
Coupled Hartree-Fock/STO-3G  &&   45.6  &&  57.0 &&  1.25 && Ref. \cite{C70:395} \\
Pople-Parr-Pariser model &&  49.4        &&  63.8         &&   1.29 && Ref. \cite{willaime:6369}] \\
Tight binding   && 77.0    &&  91.6    &&  1.19 && Ref. \cite{C70:378}\\
Bond polarizability model  && 89.2    && 109.2    && 1.22  && Ref. \cite{C70:382}\\
Valence effective Hamiltonian &&  154.0  &&214.3  &&1.39 && Ref. \cite{PhysRevB.46.16135} \\
Monopole-dipole     &&   60.8 &&   73.8 && 1.21 && Ref. \cite{C70:383} \\
MNDO/PM3  &&  63.9   &&79.0   &&1.24 && Ref. \cite{C70:393} \\
HF 6-31+G  &&  75.1  &&  89.8  &&  1.20 && Ref. \cite{jonsson:572} \\
TDDFT/SAOP && 83 && 101 && 1.22 && Ref. \cite{C70:374}\\
TDCDFT/VK && 76 && 91 && 1.51 && Ref. \cite{C70:374}\\
PBE/NRLMOL  &&         82.9  &&  102.8  && 1.24 && This work\\ 
\hline
\end{tabular}
\end{table*}

\section*{references}
\bibliographystyle{iopart-num}
\bibliography{C70_ref,list1,dunlap,list2,delta_scf,ADFT,diamondoid}

\end{document}